\documentclass{PoS}
\pdfoutput=1
\usepackage[usenames,dvipsnames,table]{xcolor}
\usepackage{graphicx}
\usepackage{color}
\usepackage{bm}
\usepackage{subfigure}
\usepackage{nameref,zref-xr}
\usepackage[mathscr]{eucal}
\usepackage{url}
 \usepackage{caption}
 
  \newcommand{\capdef}{}
  \newcommand{\mycaption}[2][\capdef]{\renewcommand{\capdef}{#2}
       \caption[#1]{{\footnotesize #2}}} 

\title{Looking for Galactic Diffuse Dark Matter \\
in INO-MagICAL Detector}

\ShortTitle{Dark Matter searches in INO-MagICAL}

\author{\speaker{Sanjib Kumar Agarwalla}\\
Institute of Physics, Sachivalaya Marg, Sainik School Post, Bhubaneswar 751005, India\\ 
Homi Bhabha National Institute, Training School Complex, Anushakti Nagar, Mumbai 400085, India\\ 
E-mail: \email{sanjib@iopb.res.in}}

\author{Amina Khatun\\
Institute of Physics, Sachivalaya Marg, Sainik School Post, Bhubaneswar 751005, India\\ 
Homi Bhabha National Institute, Training School Complex, Anushakti Nagar, Mumbai 400085, India\\ 
        E-mail: \email{amina@iopb.res.in}}

\author{Ranjan Laha  \\     
Kavli Institute for Particle Astrophysics and Cosmology (KIPAC), Department of Physics, Stanford University, Stanford, CA 94305, USA\\
SLAC National Accelerator Laboratory, Menlo Park, CA 94025, USA\\
E-mail: \email{rlaha@stanford.edu}}
	 
\abstract{The Weakly Interacting Massive Particle (WIMP) is a popular particle physics  
candidate for the dark matter (DM). It can annihilate and/or decay to neutrino and 
antineutrino pair. The proposed 50 kt Magnetized Iron CALorimeter (MagICAL) detector
at the India-based Neutrino Observatory (INO) can observe these pairs over the 
conventional atmospheric neutrino and antineutrino fluxes. If we do not see any excess
of events in ten years, then INO-Magical can place competitive limits on self-annihilation 
cross-section ($\langle\sigma v\rangle$) and decay lifetime ($\tau$) of dark matter 
at 90\% C.L.: $\langle\sigma v\rangle\leq 1.87\,\times\,10^{-24}$ cm$^3$ s$^{-1}$ and 
$\tau\geq 4.8\,\times\,10^{24}$ s for $m_\chi$ = 10 GeV assuming the NFW 
as DM density profile.} 

\FullConference{The 19th International Workshop on Neutrinos from Accelerators-NUFACT 2017\\
		25-30 September, 2017\\
		Uppsala University, Uppsala, Sweden}

\begin{document}

\section{Introduction}
\label{sec:Intro}

The existence of mysterious dark matter is confirmed by various astrophysical~\cite{Strigari:2013iaa} 
and cosmological observations~\cite{Steigman:2012ve} through the gravitational 
interaction. The data collected by Planck satellite reveal that around $26\%$ of total 
energy density of the Universe is composed of dark matter~\cite{Ade:2015xua}. 
The particle nature of dark matter is completely unknown. However, with the particle having 
$\sim$100 GeV mass and interaction strength of the order of electro-weak 
coupling, the predicted relic abundance of cold dark matter (CDM) in the 
Universe matches with its current value. We call these class of dark matter (DM) as Weakly 
Interacting Massive Particle (WIMP)~\cite{Jungman:1995df}. 
If there is non-zero coupling between dark matter and Standard Model (SM) particles, 
then we can detect dark matter by indirect way. For example, due to annihilation of
dark-matter to any SM particles, we can observe an excess of the stable particles, like photons, 
neutrinos, which are created in the final state of SM particle's decay chain. In this study, 
we assume that dark matter ($\chi$) self-annihilate to neutrino and antineutrino pair, and 
explore the phenomenological consequences in context of the  Magnetized Iron CALorimeter 
(MagICAL) detector proposed by the India-based Neutrino Observatory (INO)~\cite{INO} project.

The MagICAL is designed to detect atmospheric neutrinos having multi-GeV energy 
and coming from all possible directions. The main physics aim of this experiment 
is to determine the neutrino mass ordering using the Earth matter effect and to 
measure the atmospheric oscillation parameters precisely~\cite{Ahmed:2015jtv}. 
In this paper, we show that the MagICAL detector can also play a very important 
role to look for Galactic diffuse dark matter having mass in the multi-GeV range.

\section{Dark Matter Inputs}
\label{sec:input}

The spherically symmetric dark matter density parameterization is given by
\begin{equation}
\rho(r) = \frac{\rho_{0}}{\big[\delta\,+\,r/r_{s}\big]^\gamma\,.\,\big[1+(r/r_{s})^\alpha\big]^{(\beta-\gamma)/\alpha}} \,.
\label{eq1}
\end{equation}
The parameter $\rho(r)$ denotes the density as a function of distance $\it{r}$  
from the center of the galaxy, and $r_{s}$ is the scale radius. The shape of the profile 
is controlled by $\alpha$ and $\beta$, $\gamma$,  and $\delta$. The local dark matter 
density at the Solar radius (R$_{sc}$) is $\rho_{sc}$. We take R$_{sc}$ = 8.5 kpc.  
The parameter $\rho_{0}$ is the normalization constant. We produce all the results  
for two different DM profiles: the Navarro-Frenk-White (NFW) profile~\cite{Navarro:1995iw}, 
and the Burkert\footnote{We do not present the results with the Burkert profile in this write-up. 
However the results with the Burkert profile are shown in Ref.~\cite{Khatun:2017adx}.}
profile~\cite{1999dmap.conf..375B}, and associated parameter values 
are given in table~\ref{table0}. 
\begin{table}[htb!]
 \centering
\begin{tabular}{|c|c|c|c|}
\hline
 & ($\alpha, \beta, \gamma, \delta$ ) & $\rho_{sc}$ [GeV cm$^{-3}$] & $r_{s}$ [kpc]\\
\hline
NFW & (1, 3, 1, 0) &  0.471 &  16.1 \cr 
Burkert & (2, 3, 1, 1) & 0.487 & 9.26  \cr
\hline
\end{tabular}
\mycaption{The necessary values of parameters related to dark matter profiles are taken 
from Ref.\,\cite{Aartsen:2015xej}.}
\label{table0}
\end{table}
We assume that dark matter particle ($\chi$) and its antiparticle ($\bar\chi$) 
annihilate to produce a neutrino and an antineutrino in the final state with 
100$\%$ branching ratio:
\begin{equation}
 \chi + \bar\chi \rightarrow \nu + \bar\nu \,.
 \label{eq2}
\end{equation}
The ratio of $\nu_e$, $\nu_\mu$, and $\nu_\tau$ at the source are assumed to be 1:1:1, 
which remains unchanged after reaching  the Earth surface due to loss of coherence 
in flight over astrophysical distances.
The $\nu/ \bar{\nu}$ flux of each flavor of per unit energy range per unit solid angle 
originated from the dark matter particles annihilation is given by
\begin{equation}
 \frac{d^2\Phi^{ann}_{\nu/\bar\nu}}{{dE}\,d\Omega} = \frac{\langle \sigma_A v\rangle}{2} J^{ann}_{\Delta\Omega}\,
 \frac{R_{sc}\rho^2_{sc}}{4\pi\,m_{\chi}^2}\,\,\frac{1}{3} \frac{{dN^{ann}}}{{dE}}\,.
 \label{eq6}
\end{equation}
In above, $\langle \sigma_A v\rangle$ is the self-annihilation cross-section and $m_{\chi}$ 
is mass of DM particles.  Integrating the square of dark matter density over the whole sky and then taking 
average over $4\pi$ solid angle, we calculate $J^{ann}_{\Delta\Omega}$, which is obtained as 
3.33 for the NFW profile and 1.6 for the Burkert profile. 
The factor $\frac{1}{2}$ appears as we assume the dark matter particle is same as its 
own antiparticle. The factor $\frac{1}{3}$ takes care the flavor ratio of 
$\nu/ \bar\nu$ on the Earth's surface. For the isotropic production of $\nu$ and $\bar\nu$ at source, 4$\pi$ 
comes in the denominator. As dark matter is non-relativistic, $\nu/ \bar{\nu}$ energy 
spectrum is written as   
\begin{equation}
\frac{{dN^{ann}}}{{dE}} = \delta(E_{\nu/\bar\nu} - m_\chi)\,.
\label{eq7}
\end{equation}

In case of decay of dark matter through $\chi \rightarrow \nu + \bar\nu\,$ channel with 100$\%$ 
branching ratio (where the final state neutrino can be of any flavor), $\nu/\bar\nu$ flux can be 
written as
\begin{equation}
\frac{d^2\Phi^{dec}_{\nu/\bar\nu}}{{dE}\,d\Omega} = J^{dec}_{\Delta\Omega}\,
\frac{R_{sc}\rho_{sc}}{4\pi\,m_{\chi}\,\tau}\,\, \frac{1}{3} \frac{{dN^{dec}}}{{dE}}\,.
\label{eq10}
\end{equation}
The factor $\frac{1}{3}$ and  4$\pi$ appear due to the same reasons as described for the 
annihilation case. The parameter $J^{dec}_{\Delta\Omega}$ represents the average value of
line of sight integration of dark matter density over whole sky. We get $J^{dec}_{\Delta\Omega} 
= 2.04$ for the NFW profile and 1.85 for the Burkert profile. For decaying dark matter, the 
energy spectrum of neutrino is given by
\begin{equation}
\frac{{dN^{dec}}}{{dE}} = \delta(E_{\nu/\bar\nu} - m_\chi /2)\,.
\label{eq11}
\end{equation}

\begin{figure*}[htb!]
\[
\begin{array}{cc}
\includegraphics[width=.49\linewidth]{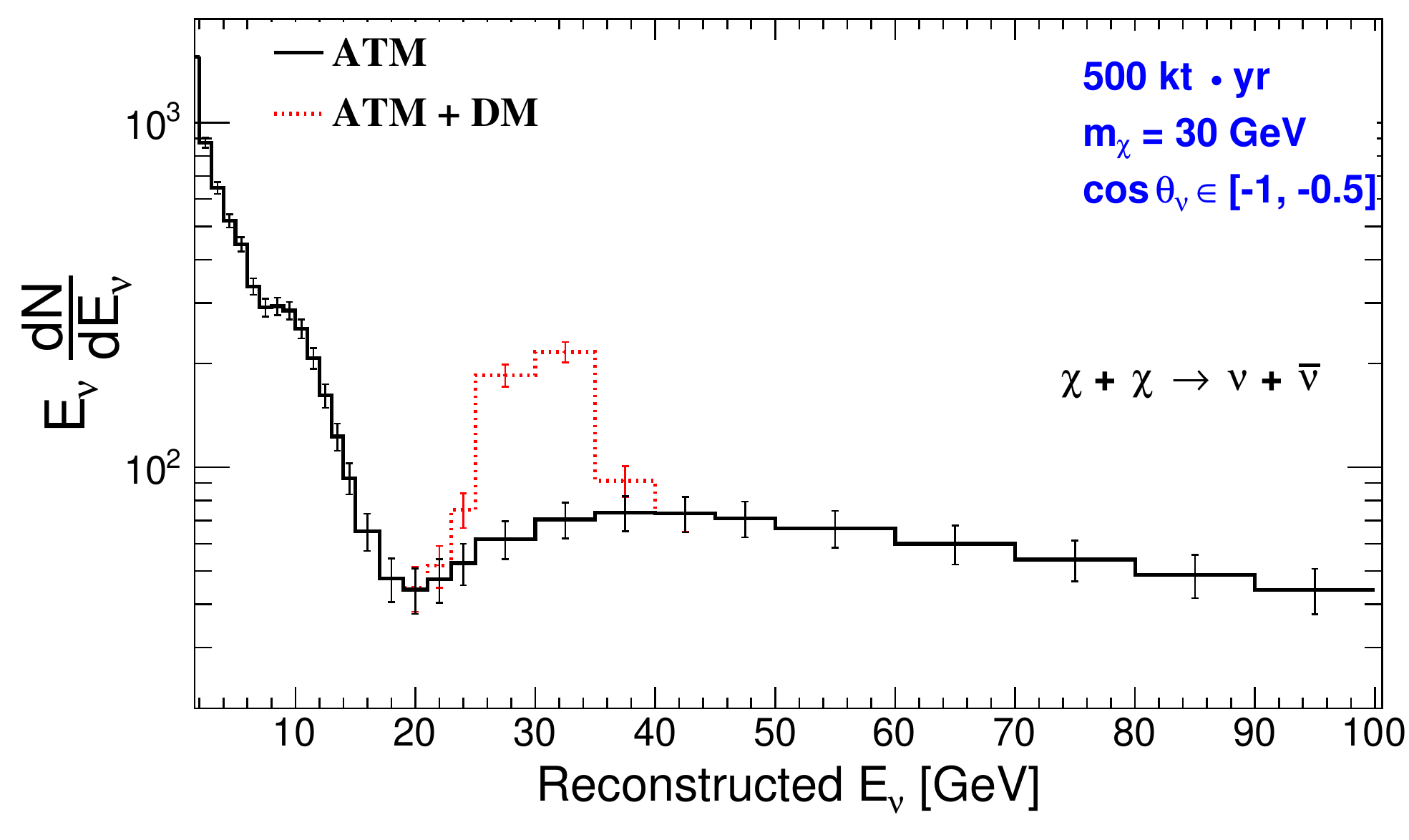}& 
\includegraphics[width=.49\textwidth]{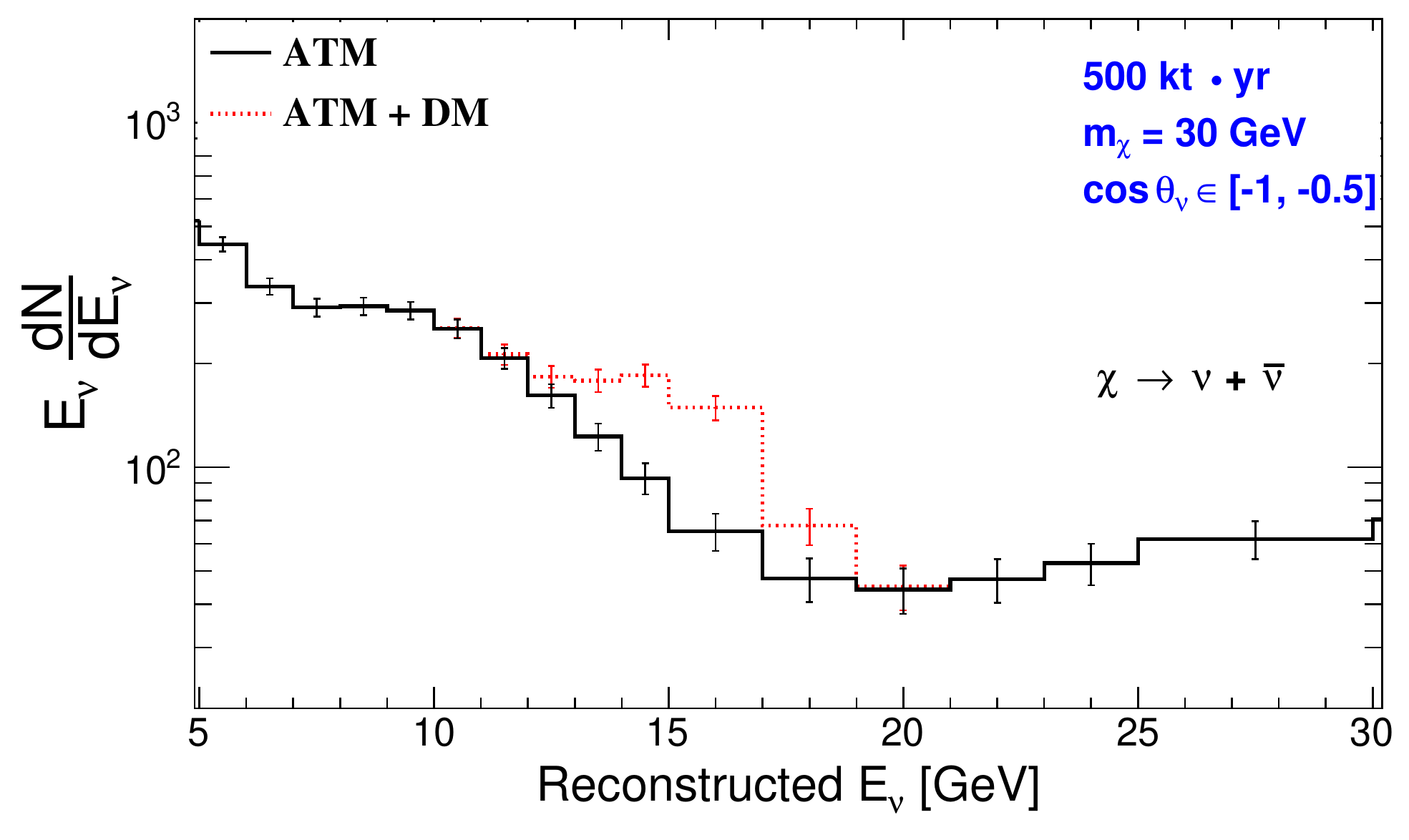}\\
\end{array}
\]
\mycaption{Event spectra of atmospheric $\nu_\mu$ is presented by black solid line. 
The red dashed lines are for events in presence of annihilation (left panel) and decay 
(right panel) of dark matter having 30 GeV mass. Both panels are for integrating over
$\cos\theta\in[-1,-0.5]$ using 500 kt$\cdot$yr exposure of MagICAL. The choice of 
mass hierarchy is normal ordering (NO).}
\label{fig4}
\end{figure*}

\section{Results } 
\label{sec:results}
This section is devoted to show results. In Fig.\,\ref{fig4}, we present the atmospheric neutrino 
events integrated over the reconstructed neutrino zenith angle $\cos\theta\in[-1,-0.5]$ using 500 kt$\cdot$yr
exposure of MagICAL. The black solid and red dashed lines represents the events in absence (ATM) and presence 
of dark matter (ATM + DM) respectively. The left panel is for dark matter annihilation and right panel is for 
decaying dark matter. Here, we assume that dark matter has the mass of 30 GeV. In this 
case, each of the final state $\nu$ and $\bar\nu$ from the dark matter annihilation (decay) have 
30 (15) GeV energy. Therefore we see an excess of $\nu_\mu$ events around 30 (15) GeV of reconstructed neutrino 
energy for annihilating (decaying) dark matter in Fig.\,\ref{fig4}. 
\begin{figure}[htb!]
\begin{center}
\subfigure{\includegraphics[width =7.5cm]{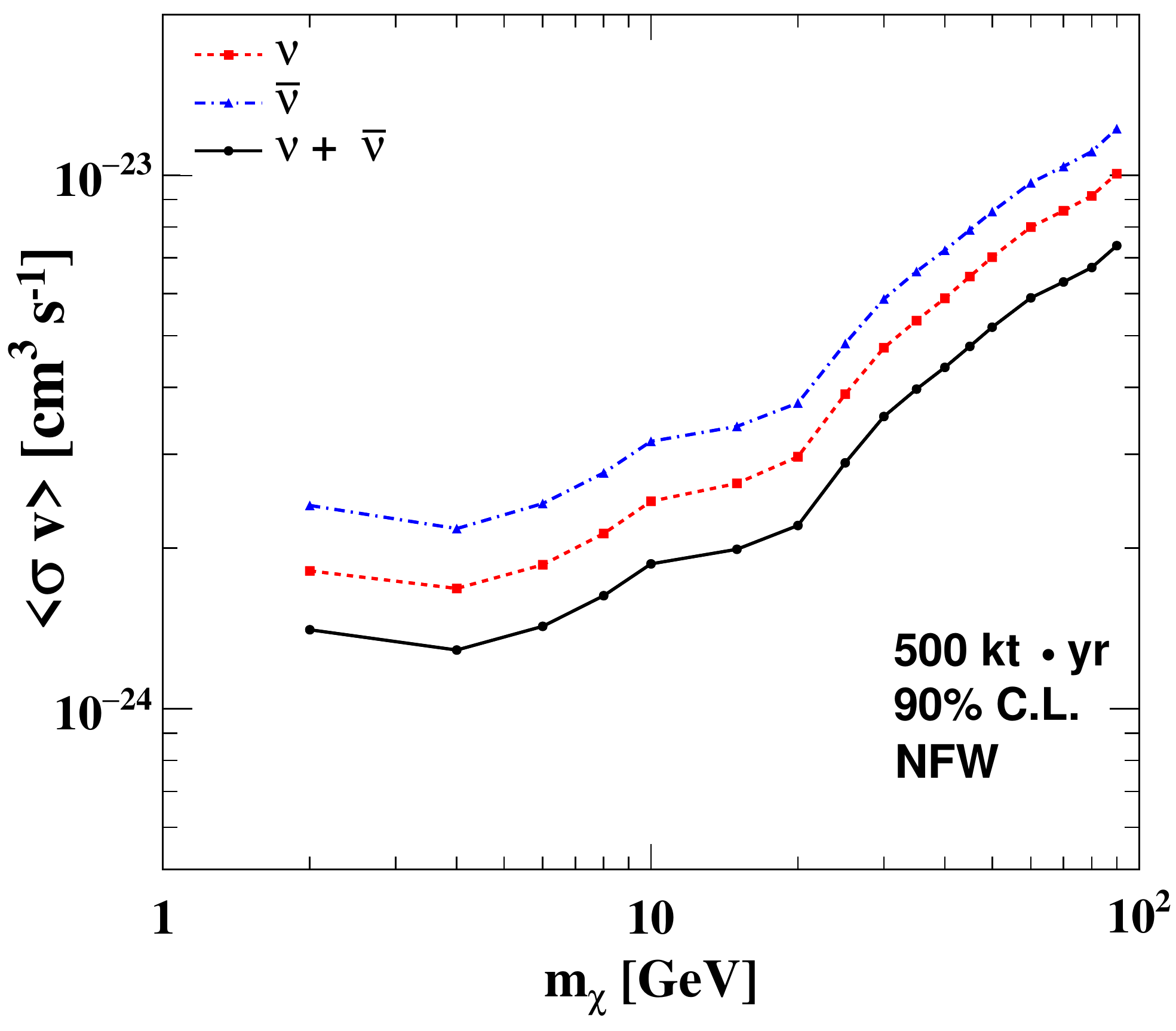}}
\subfigure{\includegraphics[width =7.5cm]{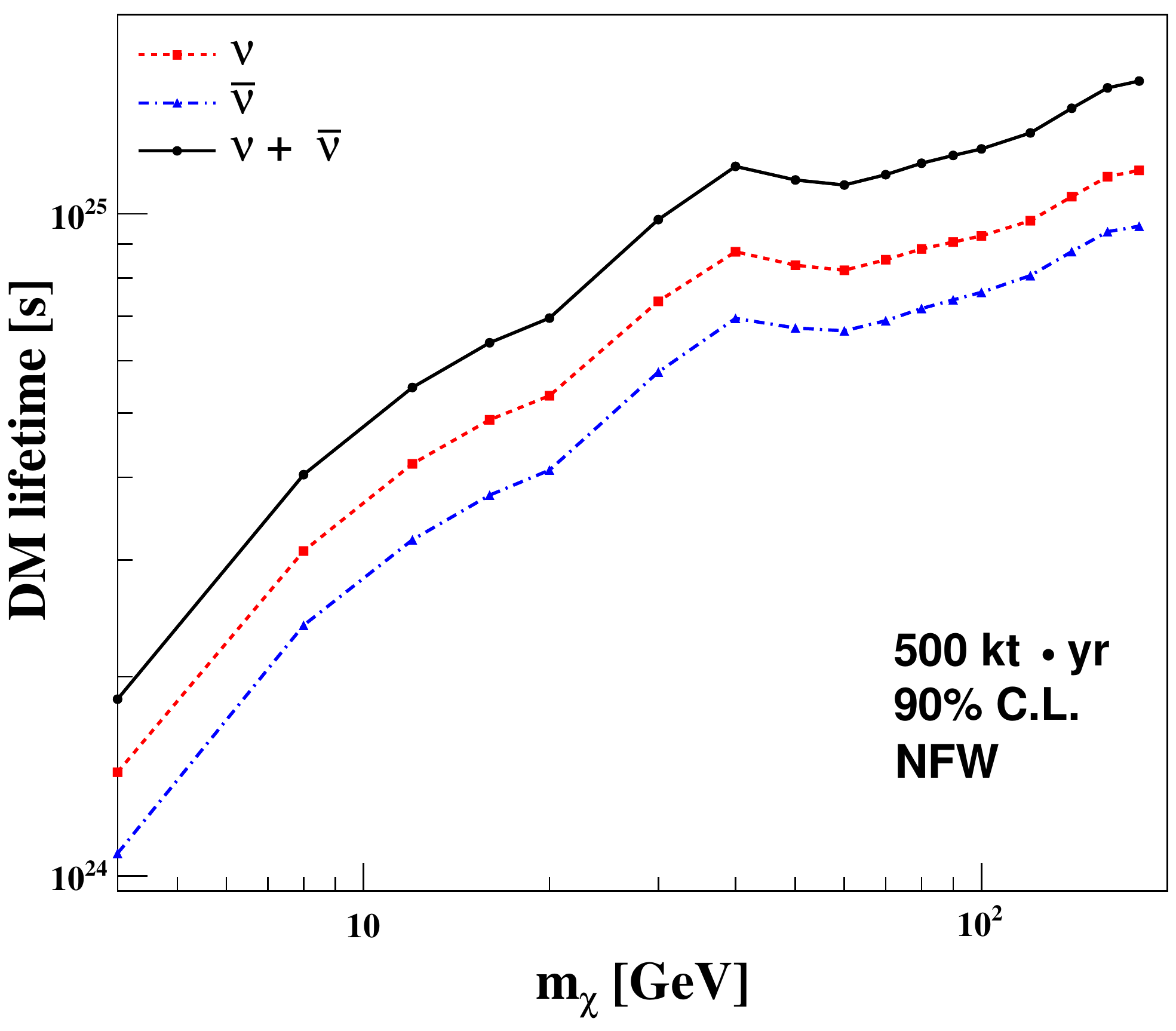}}
\mycaption{Left panel presents the upper limit on $\langle\sigma v\rangle$ 
at 90$\%$ C.L. (1 d.o.f.) for the process $\chi \chi \rightarrow \nu \bar{\nu}$. 
The right panel shows the lower bounds on decay lifetime for 
$\chi \rightarrow \nu \bar{\nu}$ at  90$\%$ C.L. (1 d.o.f.). 
For both the cases, we use 500 kt$\cdot$yr exposure of MagICAL, 
NFW profile, and normal ordering.}
\label{fig5}
\end{center}
\end{figure}
We estimate the sensitivity of MagICAL to place the upper limits 
on the self-annihilation cross-section and lower limit on the 
dark matter decay lifetime. For the details regarding the simulation 
technique and the discussion on systematic uncertainties,
please take a look at the Ref.~\cite{Khatun:2017adx}.
Fig.~\ref{fig5} presents the upper limits on 
self-annihilation cross-section of DM particles for the process 
$\chi \chi \rightarrow \nu \bar{\nu}$ (see left panel) and lower limits 
on decay lifetime for $\chi \rightarrow \nu \bar{\nu}$ 
process (see right panel) at 90$\%$ C.L. (1 d.o.f.). The red dashed, blue dot-dashed, 
and black solid lines are obtained from only $\nu_\mu$, only $\bar{\nu}_\mu$, and 
total $\nu_\mu$ and $\bar{\nu}_\mu$ respectively. 
We use 500 kt$\cdot$yr exposure of the MagICAL detector 
and normal mass ordering (NO). An important point to be noted is that the ability 
to analyze the neutrino and antineutrino data separately helps to explore the processes 
which involve lepton number violating DM. As we go to higher energies, constraints 
on the self-annihilation cross-section get deteriorated, and for decay lifetime, 
bounds get improved. We understand these features using different 
$m_\chi$ dependence of signal to background ratio in case of annihilation 
and decay of dark matter. For detailed explanation, 
see Ref.~\cite{Khatun:2017adx}. 

\subsection{Comparison with other experiments}
\begin{figure}[htb!]
\subfigure{ \includegraphics[width =7.5cm]{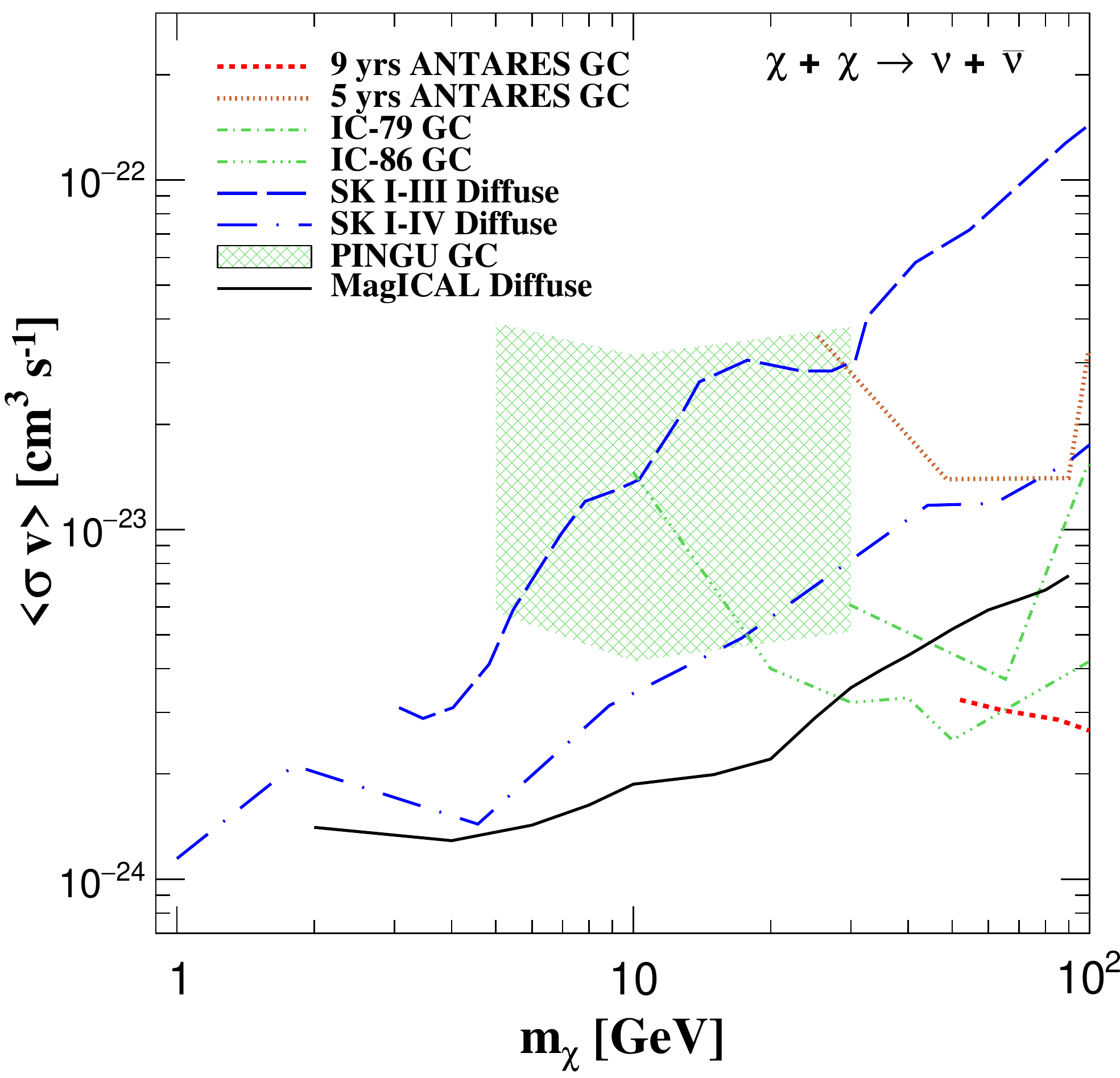}}
\subfigure{\includegraphics[width =7.5cm]{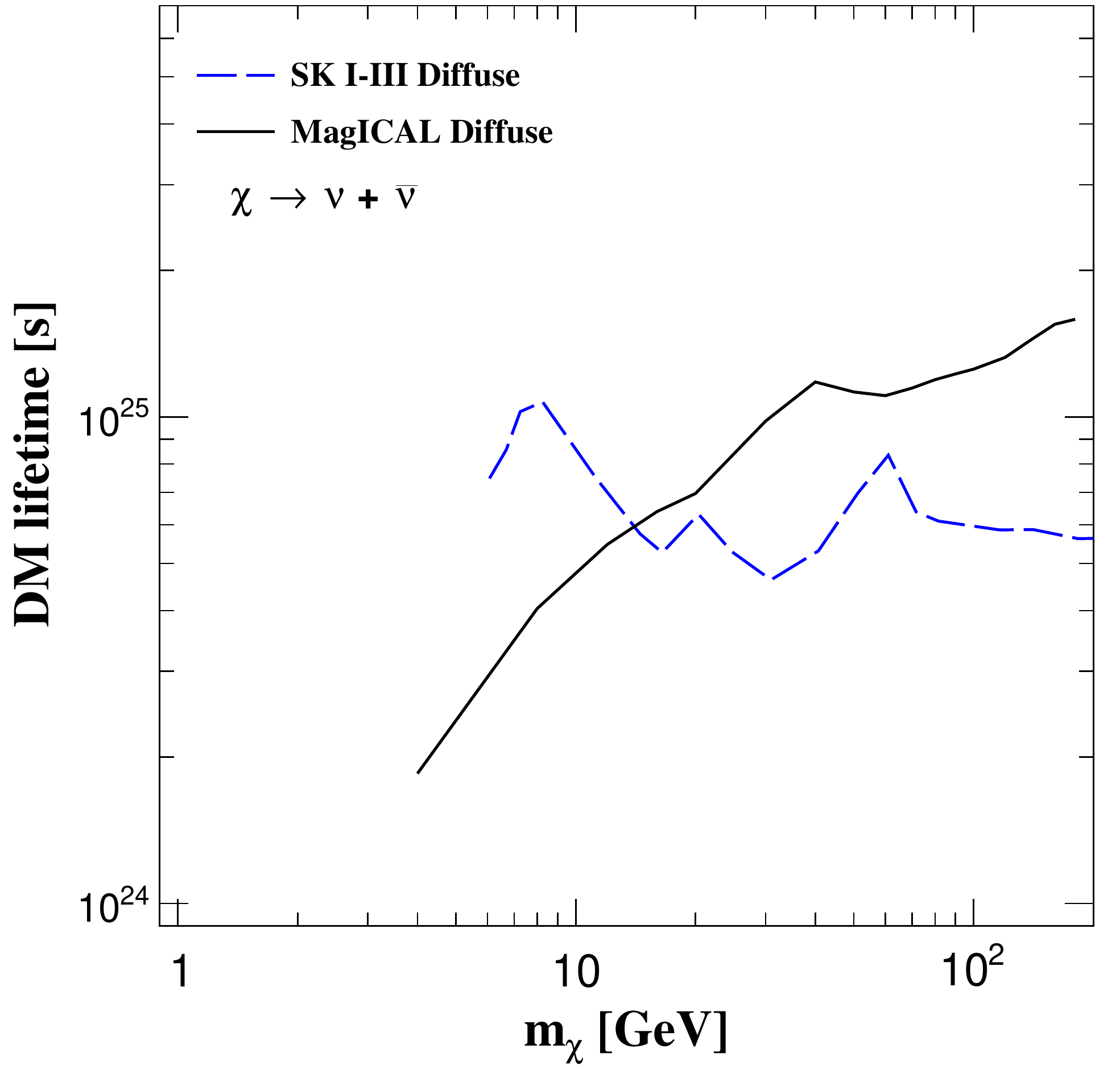}}
\mycaption{Left panel shows the current bounds on self-annihilation cross-section at 90$\%$ C.L. (1 d.o.f.) 
obtained using first three phases data of Super-Kamiokande in Ref.\,\cite{Mijakowski:2011zz} (blue long-dashed line),   
using four phases data of Super-Kamiokande in Ref.\,\cite{Mijakowski:2016cph} (blue long-dash-dotted line),   
using IceCube data in Ref.\,\cite{Aartsen:2015xej} (green dot-dashed) and Ref.\,\,\cite{Aartsen:2017ulx} 
(green triple-dot-dashed lines),  and using ANTARES data in Ref.\,\,\cite{Adrian-Martinez:2015wey} (red dotted)
and Ref.\,\cite{Albert:2016emp} (red dashed lines). The future sensitivity of PINGU\,\cite{Aartsen:2014oha} 
(green shaded region) with its one year of exposure is also shown. The limits obtained from our analysis 
using 500 kt$\cdot$yr MagICAL is plotted in black solid line. The right panel shows the current bound 
on decay lifetime of DM at 90$\%$ C.L. (1 d.o.f.) from Super-Kamiokande\,\cite{Mijakowski:2011zz} (blue long-dashed line)  
and the expected limit from MagICAL (black solid line) using 500 
kt$\cdot$yr exposure.}
\label{fig8}
\end{figure}

There are constraints on the self-annihilation cross-section from the 
experiments like Super-Kamiokande~\cite{Mijakowski:2011zz,Mijakowski:2016cph}, 
IceCube~\cite{Aartsen:2015xej,Aartsen:2017ulx}, and 
ANTARES~\cite{Adrian-Martinez:2015wey,Albert:2016emp}. In left panel of Fig.\,\ref{fig8}, 
we show these limits along with the expected bound from MagICAL with 500 kt$\cdot$yr exposure 
as obtained in this analysis (black solid line). In the multi-GeV energy range, the MagICAL detector 
is found to place the most stringent constraint than other detectors. 
We compare the current bound on the dark matter decay lifetime from the first three phases 
of Super-Kamiokande data\,\cite{Mijakowski:2011zz} and the expected limit from MagICAL detector in 
right panel of Fig.\,\ref{fig8}. 

\section{Concluding Remarks}

We study the capability of the INO-MagICAL detector to probe the Galactic diffuse 
dark matter. The expected limit on the self-annihilation cross-section of dark matter 
having mass 10 GeV is $\langle\sigma v\rangle\leq 1.87\,\times\,10^{-24}$ cm$^3$ s$^{-1}$ 
at $90\%$ C.L. (1 d.o.f.) using 500 kt$\cdot$yr exposure of MagICAL and 
assuming the NFW as dark matter density profile. The ability to distinguish the neutrino 
and antineutrino in MagICAL gives an opportunities to probe lepton number violating 
dark matter interactions. 

\bibliographystyle{JHEP}
\bibliography{reference-ical-dm.bib}

\providecommand{\href}[2]{#2}\begingroup\raggedright\begin{thebibliography}{10}

\bibitem{Strigari:2013iaa}
L.~E. Strigari, {\it {Galactic Searches for Dark Matter}},  {\em Phys. Rept.}
  {\bf 531} (2013) 1--88, [\href{http://arxiv.org/abs/1211.7090}{{\tt
  arXiv:1211.7090}}].

\bibitem{Steigman:2012ve}
G.~Steigman, {\it {Neutrinos And Big Bang Nucleosynthesis}},  {\em Adv. High
  Energy Phys.} {\bf 2012} (2012) 268321,
  [\href{http://arxiv.org/abs/1208.0032}{{\tt arXiv:1208.0032}}].

\bibitem{Ade:2015xua}
{\bf Planck} Collaboration, P.~A.~R. Ade et~al., {\it {Planck 2015 results.
  XIII. Cosmological parameters}},  \href{http://arxiv.org/abs/1502.01589}{{\tt
  arXiv:1502.01589}}.

\bibitem{Jungman:1995df}
G.~Jungman, M.~Kamionkowski, and K.~Griest, {\it {Supersymmetric dark matter}},
   {\em Phys. Rept.} {\bf 267} (1996) 195--373,
  [\href{http://arxiv.org/abs/hep-ph/9506380}{{\tt hep-ph/9506380}}].

\bibitem{INO}
India-based Neutrino Observatory (INO), http://www.ino.tifr.res.in/ino/.

\bibitem{Ahmed:2015jtv}
{\bf ICAL} Collaboration, S.~Ahmed et~al., {\it {Physics Potential of the ICAL
  detector at the India-based Neutrino Observatory (INO)}},
  \href{http://arxiv.org/abs/1505.07380}{{\tt arXiv:1505.07380}}.

\bibitem{Navarro:1995iw}
J.~F. Navarro, C.~S. Frenk, and S.~D.~M. White, {\it {The Structure of cold
  dark matter halos}},  {\em Astrophys. J.} {\bf 462} (1996) 563--575,
  [\href{http://arxiv.org/abs/astro-ph/9508025}{{\tt astro-ph/9508025}}].

\bibitem{Khatun:2017adx}
A.~Khatun, R.~Laha, and S.~K. Agarwalla, {\it {Indirect searches of Galactic
  diffuse dark matter in INO-MagICAL detector}},  {\em JHEP} {\bf 06} (2017)
  057, [\href{http://arxiv.org/abs/1703.10221}{{\tt arXiv:1703.10221}}].

\bibitem{1999dmap.conf..375B}
A.~{Burkert} and J.~{Silk}, {\it {On the structure and nature of dark matter
  halos}},  in {\em Dark matter in Astrophysics and Particle Physics} (H.~V.
  {Klapdor-Kleingrothaus} and L.~{Baudis}, eds.), p.~375, 1999.
\newblock \href{http://arxiv.org/abs/astro-ph/9904159}{{\tt astro-ph/9904159}}.

\bibitem{Aartsen:2015xej}
{\bf IceCube} Collaboration, M.~G. Aartsen et~al., {\it {Search for Dark Matter
  Annihilation in the Galactic Center with IceCube-79}},  {\em Eur. Phys. J.}
  {\bf C75} (2015), no.~10 492, [\href{http://arxiv.org/abs/1505.07259}{{\tt
  arXiv:1505.07259}}].

\bibitem{Mijakowski:2011zz}
P.~Mijakowski, {\em {Direct and Indirect Search for Dark Matter}}.
\newblock PhD thesis, Warsaw, Inst. Nucl. Studies, 2011.

\bibitem{Mijakowski:2016cph}
{\bf Super-Kamiokande} Collaboration, P.~Mijakowski, {\it {Indirect searches
  for dark matter particles at Super-Kamiokande}},  {\em J. Phys. Conf. Ser.}
  {\bf 718} (2016), no.~4 042040.

\bibitem{Aartsen:2017ulx}
{\bf IceCube} Collaboration, M.~G. Aartsen et~al., {\it {Search for Neutrinos
  from Dark Matter Self-Annihilations in the center of the Milky Way with 3
  years of IceCube/DeepCore}},  \href{http://arxiv.org/abs/1705.08103}{{\tt
  arXiv:1705.08103}}.

\bibitem{Adrian-Martinez:2015wey}
{\bf ANTARES} Collaboration, S.~Adrian-Martinez et~al., {\it {Search of Dark
  Matter Annihilation in the Galactic Centre using the ANTARES Neutrino
  Telescope}},  {\em JCAP} {\bf 1510} (2015), no.~10 068,
  [\href{http://arxiv.org/abs/1505.04866}{{\tt arXiv:1505.04866}}].

\bibitem{Albert:2016emp}
A.~Albert et~al., {\it {Results from the search for dark matter in the Milky
  Way with 9 years of data of the ANTARES neutrino telescope}},
  \href{http://arxiv.org/abs/1612.04595}{{\tt arXiv:1612.04595}}.

\bibitem{Aartsen:2014oha}
{\bf IceCube PINGU} Collaboration, M.~G. Aartsen et~al., {\it {Letter of
  Intent: The Precision IceCube Next Generation Upgrade (PINGU)}},
  \href{http://arxiv.org/abs/1401.2046}{{\tt arXiv:1401.2046}}.

\end{thebibliography}\endgroup
\end{document}